

\magnification=1200
\hoffset=.2truein \voffset=-.5in
\hsize=6.7truein\vsize=8.5truein
\parindent 10pt
\parskip 1pt plus 2pt
\baselineskip=20truept
\overfullrule=0pt
\hfill\vbox{\hbox{TIFR/TH/92-66}\hbox{IISc/CTS/92-12}\hbox{November,
1992}}

\centerline{\bf BOSONIC MEAN FIELD THEORY OF THE SPIRAL PHASES}
\centerline{\bf OF HEISENBERG ANTIFERROMAGNETS ON A CHAIN}
\vskip 0.7truein
\centerline{Sumathi Rao$^{\star\dag}$}\footnote{}{$^{\star}$ e-mail address:
SUMATHI@TIFRVAX.BITNET}
\footnote{}{$^{\dag}$ Permanent Address:Institute of Physics,
Sachivalaya Marg, Bhubaneswar 751005, India.}

\centerline{Tata Institute of Fundamental Research, Homi Bhabha Road,
Bombay 400005, India}
\centerline {and}

\centerline{Diptiman Sen$^{\ddag}$}\footnote{}{$^{\ddag}$ e-mail address:
DIPTIMAN@CTS.IISC.ERNET.IN}

\centerline{Centre for Theoretical Studies, Indian Institute of Science,
Bangalore 560012, India}

\vskip 0.7truein
\centerline{\bf ABSTRACT}

We develop a novel bosonic mean field theory to describe the spiral phases
of a Heisenberg antiferromagnet on a one-dimensional chain, in terms of
three bosons at each site. The ground state is disordered and for large
values of the spin $S$, two different and exponentially small energy gaps
are found. The spin-spin correlation function is computed and is shown to
decay exponentially at large distances. Our mean field theory is also
shown to be exact in a large-$N$ generalization.


 \def\AFF{[4]} \def\TWO{[5]}  
 \def\CHAN{[7]} \def\COL{[8]}  
\def\ARO{[11]} \def\SAR{[12]} \def\MIL{[13]} \def\DAG{[14]} \def\NUMNEEL{[15]}
\def\FERRER{[16]} \def\CHITRA{[17]} \def\MG{[18]} \def\RAO{[19]} \def\AZA{[20]}
\def\VILLAIN{[21]} \def\JAPAN{[22]}

\def\l{\lambda}
\def\bs{{\bf S}}
\def\bsn{\bs_n}
\def\bsm{\bs_m}
\def\bso{\bs_0}
\def\ok{\omega_k}
\def\muk{\mu_k}
\def\nuk{\nu_k}
\def\eabc{\epsilon_{\alpha\beta\gamma}}
\def\dab{\delta_{\alpha\beta}}
\def\ama{a_{m\alpha}}
\def\ana{a_{n\alpha}}
\def\amb{a_{m\beta}}
\def\anb{a_{n\beta}}
\def\d{\delta}
\def\t{\theta}
\def\g{\gamma}
\def\a{\alpha}
\def\b{\beta}
\def\xmn{X_{m,n}}
\def\xon{X_{0,n}}
\def\ymn{Y_{m,n}}
\def\yon{Y_{0,n}}
\def\xn{X_{n,n+1}}
\def\xnn{X_{n,n+2}}
\def\yn{Y_{n,n+1}}
\def\ynn{Y_{n,n+2}}
\def\io{I_0}
\def\ita{I_{\t/a}}
\def\e{\epsilon}
\def\D{\Delta}
\def\Do{\D_0}
\def\Dt{\D_{\t}}
\def\eqn{\eqno}
\vfil
\eject

The study of the ground state and excitation spectrum of two dimensional
quantum antiferromagnets has aroused considerable
interest [1--14],
particularly since their relevance to high $T_c$ superconductivity was
realized [1,2].  A wide variety of approaches have been used - field
theory methods [3--6], linear spin-wave theories\CHAN, analogy with
neutral superfluidity \COL, bosonic and fermionic mean field
theories [9--13], and numerical methods \DAG\ - to study the problem.

The first question that was tackled was the ground state and excitation
spectrum of unfrustrated Heisenberg antiferromagnets. Field theory methods
showed that in one dimension, the ground state and its excitations had
completely different properties depending upon whether the spin was an
integer or half-integer. For integer spins, the ground state was
exponentially  disordered at long distances and excitations had a gap,
whereas for half-integer spins, the ground state was only algebraically
disordered and had massless excitations [3,4]. This difference had its
origin in the topological term that was induced in the long wave-length
effective field theory. But this difference disappeared in two dimensions,
where no topological term was found to be induced \TWO.  Moreover,
numerical evidence \NUMNEEL\ favoured a Neel ordered ground state in two
dimensions.

The question of the ground state and excitation spectrum of a frustrated
model is far more complex and is, as yet,  not completely understood. The
mapping to conformal field theories \AFF\ has led to the expectation that
the phase diagram for half-integer spins has a region where the model has
algebraic disorder and massless excitations. Outside this region, the
system is expected to be dimerized, except at specific points. For integer
spins, the region of massless excitations is replaced by massive
excitations. However, there exist few explicit results. For two
dimensional models, the phase diagram is even more uncertain, because the
various different approaches lead to different answers. In particular, the
existence of a spin-liquid state (predicted by fermionic mean field
theories) is still unconfirmed. (For a recent summary of existing results,
see Ref. \FERRER.)

In this paper, we introduce a new (bosonic) mean field
theory (MFT) involving a representation of spins by three bosons in the
adjoint representation of the $SO(3)$ group of spins, precisely to study
this question. In particular,  we address the following specific issue.
We study the frustrated Heisenberg antiferromagnet (AFM) on a chain using
our three boson representation (3BR), with the aim of obtaining explicit
results.  This method (like all mean field methods) is insensitive to the
presence of topological terms. However, since the ultimate aim is to study
spirals in two dimensions, where topological terms are not expected, this
is not a serious handicap.  This 3BR works well for
spiral phases, reproducing the zero modes at $q=0$ and $q=\theta$ (where
$\theta$ is the spiral angle) as expected from symmetry considerations.
The same representation also works for the spiral phases of the triangular
AFM in two dimensions and is expected to work for the helicoidal phases of
the frustrated model on a two dimensional square lattice. One advantage
of the 3BR is that no rotation of all the spins to a ferromagnetic
configuration is needed. This makes an analysis of helicoidal phases in two
dimensions technically much simpler \CHITRA.

In the 3BR employed here, we represent the spins at every site by a
triplet of bosons. To enforce the spin nature of the operators, two
constraints are required at each site which is in contrast
to the more commonly
used representation of spins in terms of two Schwinger bosons (2BR) where only
one constraint per site is required. These two
constraints are imposed on an
average. Following the method of Sarker et al\SAR, we perform a
Hartree-Fock-Mean-Field (HF-MF) averaging to obtain the ground state
energy in terms of six variational parameters, whose values, in turn, are
obtained by extremizing the energy. In the $S\rightarrow\infty$ limit, the
spin-wave spectrum is reproduced. For $S$ large but finite, our solution
yields two exponentially small energy gaps. We can compute the two spin
correlation function and show that beyond some length scale, the
correlation function falls off exponentially. Finally, we show that our
HF-MF treatment becomes exact in the large-$N$ limit by generalizing the
3BR (forming the triplet representation of $SO(3)$) to an $N$BR (forming
the $N$-plet representation of $SO(N)$).

Let us start with the frustrated Heisenberg AFM on a chain described by
the Hamiltonian
\def\eone{(1)}
$$
H ~=~ \sum_n ~\left(~\bsn\cdot\bs_{n+1} ~+~ \delta ~\bsn\cdot\bs_{n+2}~
\right)
\eqn\eone
$$
where we have normalized the exchange constant $J=1$. For $\d < 1/4$, the
classical ground state is Neel ordered, whereas for $\d > 1/4$, the
classical ground state is a spiral where all the spins lie in a plane and
the relative angle $\t$ between any two spins is given by
\def\extra{(2)}
$$
{\rm cos}~\t ~=~-1/4\d
\eqn\extra
$$
so that  $\pi/2 < \t < \pi$. The classical ground state energy per
unit spin is given by
\def\etwo{(3)}
$$
{E_0\over N} ~=~ - ~(\d ~+~ {1\over 8\d})~S^2
\eqn\etwo
$$
to leading order in $S$, where $N$ is the total number of spins. For $\d =
1/2$, (a special case called the Majumdar-Ghosh model \MG\ ), this model
was recently studied by us \RAO. In the long distance, large-$S$ limit, we
mapped the model to an $SO(3)$-valued field theory and using the
$\beta$-functions of the theory, we showed that the ground state was
exponentially disordered and exhibited a gap. However, we were unable to
generalize that method to arbitrary $\delta$. (See also Ref. \AZA\ ).

We first perform a spin-wave analysis of this Hamiltonian using
Villain's action-angle variables \VILLAIN. For the general spiral case
- $i.e.$, $\delta > 1/4$, - the spin-wave spectrum \JAPAN\ (valid for
large-$S$) is given by
\def\ethree{(4)}
$$\eqalign{
\ok = 2S~[~&(~-~{\rm cos}\theta ~-~ \delta~{\rm cos}2\theta ~+~
{\rm cos}ka ~+~\delta~{\rm cos}2ka) \cr
&(~-~{\rm cos}\theta ~-~\delta~{\rm cos}2\theta ~+~{\rm cos}\theta~
{\rm cos}ka ~+~\d~{\rm cos}2\t~{\rm cos}2ka)~]^{1/2}.}
\eqn\ethree
$$
(The lattice spacing is $a$). Within the first Brillouin zone, $\ok$
vanishes at $ka=0,+\t$ and $-\t$
with a linear dispersion, and the spin-wave velocities are given by
\def\efour{(5)}
$$\eqalign{
c_0 &~=~ Sa~(1+4\d)~(1-1/16\d^2)^{1/2}\cr
{\rm and}\quad
c_{\t} &~=~c_0~(1 - 1/2\d +1/8\d^2)^{1/2}}
\eqn\efour
$$
at $ka=0$ and $ka=\pm\t$ respectively.
Thus, the mode at $k=0$ has a higher velocity than the two modes at $ka
=\pm\t$. We shall see that this spectrum is reproduced by our bosonic MFT in
the large-$S$ limit.

We now set up the bosonic MFT by expressing the spin in terms of bosons.
The 3BR expresses the components of a spin $\bs$ in terms of three bosons
as
\def\efive{(6)}
$$
S_{\a}~=~ -i~\eabc ~a_{\b}^{\dagger} ~a_{\g}
\eqn\efive
$$
where $\a,\b$ and $\g$ run from 1 to 3, $\eabc$ is completely
antisymmetric and repeated indices are always summed over.
Using $[a_{\a},a_{\b}^{\dagger} ~]=\dab$, we can check that the spin algebra is
satisfied. But to also have $S_{\a}S_{\a} = S^2 = S(S+1)$, we need to
impose the constraints
\def\esix{(7)}
$$
a_{\a}^{\dagger}~ a_{\a} ~=~ S  \quad
{\rm and} \quad a_{\a}^{\dagger}~a_{\a}^{\dagger} ~a_{\b} ~a_{\b} ~=~ 0
\eqn\esix
$$
on any physical state. Notice that the first equation in Eq. \esix\ implies
that the 3BR works only for integer spins. By enforcing these constraints,
we can check that  for any $S$, the total number of orthonormal states is
$2S+1$ as expected. The second constraint in Eq. (6) can equivalently be
rephrased as the two constraints
\def\esixc{(8)}
$$
a_{\a}^{\dagger}~a_{\a}^{\dagger} ~=~0 \quad {\rm and} \quad
a_{\b}~a_{\b}~=~0
\eqn\esixc
$$
in the sense of matrix elements between any two physical states. This
is the form in which it is employed later.

To understand the connection between spin order parameters and appropriate
expectation values of the bosons, consider a spin operator lying in the
${\hat x}$-${\hat y}$ plane, in the $S \rightarrow \infty$ limit, - $i.e.$,
in the classical limit, - with expectation value
\def\eseven{(9)}
$$
<S_{\a}> ~=~ S~({\rm cos}\phi,~ {\rm sin}\phi,~ 0).
\eqn\eseven
$$
Then Eqs. (5) and (6) imply that
\def\eeight{(10)}
$$
<a_1> ~=~ i ~{\sqrt{S\over 2}}~{\rm sin}\phi, \quad <a_2> ~=~
-i ~{\sqrt{S\over 2}}~{\rm cos}\phi, \quad <a_3> ~=~ {\sqrt{S\over 2}}
\eqn\eeight
$$
upto an arbitrary phase multiplying all the bosons. So if $\phi_{mn}$ is
the angle between the $m^{th}$ and $n^{th}$ spins, we have
\def\enine{(11)}
$$
<\bsm\cdot\bsn> ~\simeq ~<\bsm>~ \cdot ~<\bsn> ~=~ S^2~ {\rm cos}\phi_{mn}
\eqn\enine
$$
which in turn implies that
\def\eten{(12)}
$$\eqalign{
<\ama^{\dagger}\ana> ~&\simeq~<\ama^{\dagger}>~<\ana> ~=~ S~{\rm cos}^2 ~
(\phi_{mn}/2) \cr
<\ama\ana> ~&\simeq~ <\ama>~<\ana> ~=~ S~{\rm sin}^2 ~ (\phi_{mn}/2),}
\eqn\eten
$$
again upto arbitrary overall phases.
Thus, spiral ordering of spins implies non-zero expectation values of
bosonic bilinears. In fact, $<\ama^{\dagger}\ana>$ is the Ferromagnetic (FM)
order parameter and  $<\ama\ana>$ is the Antiferromagnetic (AFM)
order parameter.

Next, we observe that  the product $\bsm\cdot\bsn$
can be written as
\def\eeleven{(13)}
$$
\bsm\cdot\bsn ~=~ :\ymn^{\dagger}~\ymn: ~-~ \xmn^{\dagger}~\xmn
\eqn\eeleven
$$
where
\def\etwelve{(14)}
$$
\ymn ~=~ \ama^{\dagger}~\ana \quad {\rm and} \quad
\xmn ~=~ \ama~\ana.
\eqn\etwelve
$$
In terms of the bilinears $\xmn$ and $\ymn$, the Hamiltonian in Eq. (1) can
be rewritten as
\def\ethirteen{(15)}
$$\eqalign{
H ~=~ \sum_n ~&[~:Y_{n,n+1}^{\dagger}~Y_{n,n+1}: ~-~
X_{n,n+1}^{\dagger}~X_{n,n+1} ~+~ \delta ~:Y_{n,n+2}^{\dagger}~Y_{n,n+2}: ~-~
\delta ~X_{n,n+2}^{\dagger}~X_{n,n+2} \cr ~&+~\lambda_n~
(\ana^{\dagger}~\ana ~-~S) ~
-~\rho_n~(\ana\ana) ~-~ \rho_n^*~\ana^{\dagger}~\ana^{\dagger}~],}
\eqn\ethirteen
$$
with $\lambda_n$,$\rho_n$ and $\rho_n^*$ being the Lagrange multiplier
fields introduced to enforce the constraints in Eqs. \esix\ and \esixc\
at each site. We now make a HF decomposition by writing
\def\efourteen{(16)}
$$
A^{\dagger}~A ~=~ <A^{\dagger}~>A ~+~ A^{\dagger}~<A> ~-~ <A^{\dagger}~>
<A>,
\eqn\efourteen
$$
where
$A=\xn,\yn,\xnn$ ~and $\ynn$ ~in turn. Such a decomposition (in contrast
to the Peierls variational decomposition, which allows for all possible
decouplings  where each four boson term is written as products of pairs in
three different ways) is justified later by a large-$N$ argument.
Further, we make the MF ansatz that the parameters $\l_n = \l$, $\rho_n =
\rho_n^* = \rho$, $<\xn>$ = $X_1$, $<\xnn>$ = $X_2$,
$<\yn>$ = $Y_1$ and $<\ynn>$ = $Y_2$ ~are all independent of $n$ and real.
Notice that this ansatz - in particular, the reality of $\rho$, $X_i$ and
$Y_i$ - breaks the local gauge invariance $\ana \rightarrow
e^{i\t_n}~\ana$ of the Hamiltonian. However, physical quantities such as
spin-spin correlations remain gauge-invariant \SAR. Also note that non-zero
values for $X_1$, $X_2$ and $Y_1$, $Y_2$ imply the existence of
short-range AFM and FM orderings respectively.

We now diagonalize the MF Hamiltonian by a Bogoliubov transformation to
obtain
\def\efifteen{(17)}
$$\eqalign{
{H_{MF} \over N} ~=~ &-~\l S ~+~ X_1^2 ~+~\d X_2^2 ~-~Y_1^2 ~-~\d
Y_2^2  \cr ~&+~
a ~\int_{0}^{\pi/a} ~ {dk \over \pi}~(~\ok~b_{k\a}^{\dagger}~b_{ka} ~+~
{3\over 2}~\ok ~-~ {3\over 2}~\muk)}
\eqn\efifteen
$$
where
\def\esixteen{(18)}
$$\eqalign{
\ok ~&=~(~\muk^2 ~-~\nuk^2~)^{1/2} \cr
\muk ~&=~ \l ~+~ 2 Y_1 ~{\rm cos}ka ~+~ 2 \d ~Y_2 ~{\rm cos} 2ka  \cr
{\rm and} \quad \nuk ~&=~ \rho ~+~ 2 X_1 ~{\rm cos}ka ~+~
2 \d ~X_2 ~{\rm cos}2ka, }
\eqn\esixteen
$$
and the bosons $b_{k\a}$ are related to the bosons $a_{k\a}$ by the
standard Bogoliubov transformation given by
\def\eseventeen{(19)}
$$\eqalign{
a_{k\a} ~&=~ {\rm cosh}\t_k ~~b_{k\a} ~+~ {\rm sinh}\t_k ~~
b_{-k\a}^{\dagger} \cr
a_{k\a}^{\dagger} ~&=~ {\rm sinh}\t_k ~~b_{-k\a} ~+~ {\rm cosh}\t_k ~~
b_{k\a}^{\dagger}. }
\eqn\eseventeen
$$
The factors of three in Eq. \efifteen\ arise because the constraints in
Eq. \esix\ have only been imposed on the average, resulting in the
decoupling of the three bosons. (This tripling of number of branches of
the spin-wave spectrum is an
unfortunate feature of this MFT. A similar doubling of number of branches
occured in the MFT based on the 2BR \ARO\SAR.) Thus, we obtain the MF
ground state energy as
\def\eeighteen{(20)}
$$
{E_{MF} \over N} ~=~ -~(~S+3/2)~\l ~+~ X_1^2 ~+~\d X_2^2 ~-~
Y_1^2 ~-~\d Y_2^2 ~+~ {3a\over 2} ~\int_{0}^{\pi/a} ~ {dk \over
\pi}~~\ok.
\eqn\eeighteen
$$
The equations for the six variational parameters $\l$, $\rho$, $X_i$ and
$Y_i$ are obtained by extremizing the energy and are given by
\def\enineteen{(21)}
$$\eqalign{
S+{3\over 2}~&=~{3a\over 2}\int_0^{\pi/a}~{dk\over \pi}~~{\muk\over\ok}\cr
0~&=~{3a\over 2}\int_0^{\pi/a}~{dk\over \pi}~~{\nuk\over\ok}\cr
Y_1~&=~{3a\over 2}\int_0^{\pi/a}~{dk\over \pi}~~{\muk\over\ok}~~
{\rm cos}~ka \cr
X_1~&=~{3a\over 2}\int_0^{\pi/a}~{dk\over \pi}~~{\nuk\over\ok}~~
{\rm cos}~ka \cr
Y_2~&=~{3a\over 2}\int_0^{\pi/a}~{dk\over \pi}~~{\muk\over\ok}~~
{\rm cos}~2ka \cr   {\rm and} \quad
X_2~&=~{3a\over 2}\int_0^{\pi/a}~{dk\over \pi}~~{\nuk\over\ok}~~
{\rm cos}~2ka. }
\eqn\enineteen
$$
These equations look rather intractable. But, in fact, it is possible to
obtain the solutions to leading order in $S$. We know that as
$S\rightarrow\infty$, the solution should approach the classical spiral
ground state configuration asymptotically. Hence, to leading order in $S$,
we must have
\def\etwenty{(22)}
$$\eqalign{
Y_1 ~=~ S~{\rm cos}^2 ~\t/2, &\quad X_1 ~=~ S~{\rm sin}^2 ~\t/2 \cr
Y_2 ~=~ S~{\rm cos}^2 ~\t, &\quad X_1 ~=~ S~{\rm sin}^2 ~\t \cr
\l ~=~ -~2S~({\rm cos}\t ~+~&\d~{\rm cos}~2\t~) \quad {\rm and}
\quad \rho ~=~ 0,}
\eqn\etwenty
$$
where $\t$ is defined in Eq. \extra.
(One can check that these are the correct asymptotic values of the
parameters by comparing them with the values given in Eq. \eten\ as well as
by comparing the dispersions given in Eqs. \esixteen\ and \ethree.
Note also that by substituting this solution in Eq. \eeighteen, the MF
energy agrees with the classical energy to leading order in $S$.)
However, with these asymptotic solutions, notice that $\ok \rightarrow 0$
at $k=0$ and $k=\t/a$. Hence, the right hand sides (R.H.S.) of
Eq. \enineteen\ are  log divergent and can only equal the left hand sides
(L.H.S.), which are large but finite, if we allow for  a small mass
generation. Let us assume that near $k\sim 0$ and $k\sim\t$, the
dispersions of $\ok$ are given by
\def\etone{(23)}
$$\eqalign{
\ok ~\simeq~ {\sqrt{\Do^2 ~+~c_0^2 ~k^2}}, \quad k~&\sim~0  \cr
{\rm and} \quad \ok ~\simeq~ {\sqrt{\Dt^2 ~+~c_{\t}^2 ~(k-\t/a)^2}},
&\quad k~\sim~\t/a,}
\eqn\etone
$$
where $c_0$ and $c_{\t}$ are the spin-wave velocities given in Eq. \efour,
and $\Do$ and $\Dt$ are the two small masses generated. In fact,
we will show that the $\D_i$ are exponentially small - $i.e.$, of
$O(e^{-S})$. Notice also that the non-singular regions on the R.H.S. of
Eq. \enineteen\ are of $O(1)$ (not of $O(S)$ ~) and do not contribute to
establishing the equality of the L.H.S. and R.H.S to $O(S)$. Hence, to
leading order in $S$, we can simply assume that the
integrals on the R.H.S. are dominated by their values at $k=0$ and
$k=\t/a$. In this limit, Eqs. \enineteen\ reduce to
\def\ettwo{(24)}
$$\eqalign{
S+3/2 ~&=~ \mu_0 ~\io ~+~ \mu_{\t/a}~\ita \cr
0 ~&=~ \nu_0 ~\io ~+~ \nu_{\t/a}~\ita  \cr
Y_1 ~&=~ \mu_0 ~\io ~+~ \mu_{\t/a}~{\rm cos}\t~\ita \cr
X_1 ~&=~ \nu_0 ~\io ~+~ \nu_{\t/a}~{\rm cos}\t~\ita  \cr
Y_2 ~&=~ \mu_0 ~\io ~+~ \mu_{\t/a}~{\rm cos}2\t~\ita \cr {\rm and} \quad
X_2 ~&=~ \nu_0 ~\io ~+~ \nu_{\t/a}~{\rm cos}2\t~\ita,  }
\eqn\ettwo
$$
with
\def\etthree{(25)}
$$\eqalign{
\io ~&=~ {3a\over 2} ~\int_0^{\e}~{dk\over \pi}~{1\over\ok} ~\simeq~
{3a\over 2\pi c_0}~~{\rm ln}~{\e c_0\over \Do} \cr {\rm and} \quad
\ita ~&=~ {3a\over 2}~\int_{\t/a-\e/2}^{\t/a+\e/2}~{dk\over \pi}~
{1\over\ok} ~\simeq~ {3a\over 2\pi c_{\t}}~~{\rm ln}~{\e c_{\t}\over \Dt}.}
\eqn\etthree
$$
where $\D /c << \e << \pi /a ~$. We can now explicitly check that
the $O(S)$ terms on both sides of
Eq. \ettwo\ are satisfied when
\def\etfour{(26)}
$$\eqalign{
\Do ~&\sim~ {c_0 \over a}~ {\rm exp}~ [-{2\pi S\over 3}~ {\sqrt {{4\d - 1}\over
{4\d+1}}}~~~] \cr {\rm and} \quad
\Dt ~&\sim~ {c_{\t} \over a}~ {\rm exp}~ [-{\pi S\over 3}~ {\sqrt {{4\d -
1}\over (4\d+1)(1-1/2\d+1/8\d^2)}}~~~] .}
\eqn\etfour
$$
To summarize, the solutions to the Eqs. \enineteen\ are given in
Eqs. \etwenty\ and they lead to two exponentially small mass gaps in the
theory given in Eqs. \etfour.

It is interesting to compare the values for the gaps in Eqs. \etfour\
for $\d=1/2$ with the values obtained for the same $\d$ using the field
theory approach \RAO. The one-loop $\beta$-function of the field theory of
the Majumdar-Ghosh model led to the single mass gap $\D\sim{\rm
exp}(-1.8S)$, whereas here, the bosonic MF treatment yields $\Do\sim~{\rm
exp}(-1.2S)$ and $\Dt\sim~{\rm exp}(-.86S)$, which agree, at least upto
the order of magnitude of coefficient of $S$. The two mass gaps obtained
here (like the two spin-waves) appear to reflect the fact that the spiral
ordering of the ground state picks a particular plane. Thus, it is not
surprising that fluctuations, and hence, onset of disorder, within the
plane and perpendicular  to the plane, have different mass scales. The
field theory method, presumably, was not sensitive enough to see this
feature.

Let us now calculate the spin-spin correlation function within the bosonic
MFT. From Eqs. \eeleven\ and \etwelve, we see that the product of any two
spins can be written as a product of bilinears, so that the spin-spin
correlation function is given by
\def\etfive{(27)}
$$
<\bso \cdot \bsn> ~=~ <:\yon^{\dagger}~\yon:> ~-~ <\xon^{\dagger} ~\xon>.
\eqn\etfive
$$
Using only the Wick contractions allowed by the HF decomposition in
Eq. \efourteen, we find that the spin-spin correlation can be written as
\def\etsix{(28)}
$$
<\bso\cdot\bsn> ~=~ |<\yon>|^2 ~-~|<\xon>|^2,
\eqn\etsix
$$
where $< \yon >$  and $< \xon >$ are obtained by using the Bogoliubov
transformation in Eq. \eseventeen\ as
\def\etseven{(29)}
$$\eqalign{
<\yon> ~&=~{3a\over 2}\int_0^{\pi/a}~{dk\over \pi}~~({\muk\over\ok}~-~1)~~
{\rm cos}~nka \cr
{\rm and} \quad <\xon> ~&=~{3a\over 2}\int_0^{\pi/a}~{dk\over \pi}~~
{\nuk\over\ok}~~ {\rm cos}~nka. }
\eqn\etseven
$$
Once again, the integrals are dominated by the regions
near $k\sim 0$ and $k\sim\t$. We now explicitly compute the correlation
function in two limiting cases. When $na$, the distance
between the two spins measured in terms of the lattice spacing $a$, is
small, - $i.e.$, $na << \Do^{-1},~\Dt^{-1}$, we find that $\yon ~\sim~
S{\rm cos}^2~(n\t/2)$ and $\xon ~\sim~{\rm sin}^2~(n\t/2)$ so that
\def\eteight{(30)}
$$
<\bso\cdot\bsn> ~=~ S^2~{\rm cos}~n\t, \quad {\rm for} \quad na << \Do^{-1},
\Dt^{-1}.
\eqn\eteight
$$
This is not surprising, because at short distances, we expect the system
to be ordered. But at long distances, - $i.e.$, when $na >> \Do^{-1},~
\Dt^{-1}$, - we find that
\def\etnine{(31)}
$$\eqalign{
<\yon>~+~<\xon> ~&\sim~ S ~\int_0^{\e}~{dk\over \pi} ~~{{\rm cos}~nka
\over {\sqrt{\mu_k - \nu_k}}} ~\sim~~ S ~e^{-na \Do/c_0}~, {\rm and} \cr
<\yon>~-~<\xon> ~&\sim~ S ~\int_{\t /a - \e /2}^{\t /a + \e /2}~
{dk\over \pi} ~~{{\rm cos}~nka
\over {\sqrt{\mu_k + \nu_k}}} ~\sim~~ S ~
{\rm cos}~n\t~ e^{-na \Dt/c_{\t}}  }
\eqn\etnine
$$
Hence, for large enough distances,
\def\ethirty{(32)}
$$
<\bso\cdot\bsn> ~\sim~ S^2~{\rm cos}n\t~~{\rm exp}~[-na({\Do\over
c_0}+{\Dt\over c_{\t}})],\quad {\rm for} \quad na >> \Do^{-1}, \Dt^{-1},
\eqn\ethirty
$$
- $i.e$, the correlation function falls off exponentially.

Let us now justify the HF decompositions or Wick contractions used in
Eqs. \efourteen\ and \etfive. Naively, the four boson terms that appear
in the product of two spins can be decomposed (or contracted) in three
different ways. However, in our MF treatment, we have only allowed one
possible decomposition. For example,
$\xmn^{\dagger}\xmn ~=~ \ama^{\dagger}~\ana^{\dagger}~\amb ~\anb$  is only
decomposed as $<\ama^{\dagger}\ana^{\dagger}~> \amb\anb ~+~
\ama^{\dagger}\ana^{\dagger}~ <\amb\anb>$. The other possible contractions
$<\ama^{\dagger}~\anb>$ and $<\ama^{\dagger}~\amb>$ which are down by
factors of 1/3, because the $SO(3)$ indices $\a$ and $\b$ are not summed
over, have been ignored. The justification for this treatment comes from
the large-$N$ generalization of the model. The 3BR is generalized to an
$N$BR ($N$ boson representation) of $SO(N)$ `spins'. In the
$N\rightarrow\infty$ limit, the other contractions which are down by a
factor of $1/N$ can certainly be ignored.

Our large-$N$ generalization is similar in spirit to the generalization of
the 2BR of $SU(2)$ spins to the $N$ boson representation of $SU(N)$ spins
discussed in Ref. \ARO. But just for completeness, we mention some details of
our large-$N$ model. We write the components of an $SO(N)$ spin $S_{\a}$
as
\def\ethone{(33)}
$$
S_{\a} ~=~ -i~g^{\a}_{\b\g}~a_{\b}^{\dagger}~a_{\g}
\eqn\ethone
$$
with $\a=1,\dots N(N-1)/2$ and $\b,\g = 1,\dots N$. Furthermore, we have
the relations
\def\ethtwo{(34)}
$$
g^{\a}_{\b\g}~=~-~g^{\a}_{\g\b} \quad {\rm and} \quad g^{\a}_{\b\g}~
g^{\a}_{\delta\e}~=~
\delta_{\b\delta}\delta_{\g\e} ~-~ \delta_{\b\e}\delta_{\g\delta}.
\eqn\ethtwo
$$
To reproduce the spin algebra and the correct number of states, the
constraints in Eq. \esixteen\ are now replaced by
\def\eththree{(35)}
$$
a_{\a}^{\dagger}a_{\a} ~=~ NS/3 \quad {\rm and} \quad
a_{\a}^{\dagger}a_{\a}^{\dagger} a_{\b} a_{\b} ~=~ 0   .
\eqn\eththree
$$
Clearly, this representation works only if $NS/3$ is an integer. The
Hamiltonian for general $N$ given by
\def\ethfour{(36)}
$$
H ~=~ {3\over N} ~\sum_n ~\left(~\bsn\cdot\bs_{n+1} ~+~
\delta~ \bsn\cdot\bs_{n+2}~\right)
\eqn\ethfour
$$
can be written in terms of the bosons in Eq. \ethone\ along with the
Lagrange multiplier fields to enforce the constraints just as was done
earlier. In fact, the entire analysis can be reproduced. Here,
however, our aim in introducing the large-$N$ formalism was only to
justify the HF decoupling procedure that we used.

To conclude, a notable feature of our analysis  is that we do not
need to rotate the spins of the ground state of interest in order to make
it look FM and then proceed with the analysis. For the 2BR, such a
rotation is usually performed \ARO-\MIL. We are currently using
the 3BR to study frustrated spin models in two dimensions \CHITRA.

\noindent {\bf Acknowledgement}
We would like to thank D. M. Gaitonde for several useful discussions.
\vfil
\eject
\centerline{\bf REFERENCES}

\item{\bf 1.}
P. W. Anderson, Science {\bf 235}, 1196 (1987).

\item{\bf 2.}
S. Chakravarty, B. I. Halperin and D. R. Nelson, Phys. Rev.
{\bf B39}, 2344 (1989).

\item{\bf 3.}
F. D. M. Haldane, Phys. Rev. Lett. {\bf 50}, 1153 (1983);
     Phys. Lett.{\bf 93A}, 464 (1983).

\item{\bf 4.}
I. Affleck, in {\it Fields, Strings and Critical Phenomena},
Les Houches, 1988, ed. E. Brezin and J. Zinn-Justin, (North Holland,
Amsterdam,1989); I. Affleck, Nucl. Phys. {\bf B265}, 409 (1986);
I. Affleck, J. Phys. Cond. Matt.{\bf 1}, 3047 (1989); R. Shankar and N. Read,
Nucl. Phys.  {\bf B 336}, 457 (1990).

\item{\bf 5.}
T. Dombre and N. Read, Phys. Rev. {\bf B38}, 7181 (1988);
E. Fradkin and M. Stone, Phys. Rev. {\bf B38}, 7215 (1988);
X. G. Wen and A. Zee, Phys. Rev. Lett. {\bf 61}, 1025 (1988);
F. D. M. Haldane, Phys. Rev. Lett. {\bf 61}, 1029 (1988).

\item{\bf 6.}
L. B. Ioffe and A. I. Larkin, Int. J. Mod. Phys. {\bf B2}, 203 (1988).

\item{\bf 7.}
P. Chandra and B. Doucot, Phys. Rev. {\bf B38}, 9335 (1988).

\item{\bf 8.}
P. Chandra, P. Coleman and A. I. Larkin, J. Phys. Cond. Matt. {\bf 2},
7933 (1990).

\item{\bf 9.}
I. Affleck and J. B. Marston, Phys. Rev. {\bf B37}, 3744 (1988);
J. B. Marston and I. Affleck, Phys. Rev. {\bf B39}, 11538 (1989).

\item{\bf 10.}
X. G. Wen, F. Wilczek and A. Zee, Phys. Rev. {\bf B39}, 11413 (1988).

\item{\bf 11.}
D. P. Arovas and A. Auerbach, Phys. Rev. {\bf B38}, 316 (1988);
A. Auerbach and D. P. Arovas, Phys. Rev. Lett. {\bf 61}, 617 (1988).

\item{\bf 12.}
S. Sarker, C. Jayaprakash, H. R. Krishnamurthy and M. Ma, Phys. Rev. {\bf
B40}, 5028 (1989).

\item{\bf 13.}
F. Mila, D. Poilblanc and C. Bruder, Phys. Rev. {\bf B43}, 7891 (1991).

\item{\bf 14.}
A. Moreo, E. Dagotto, Th. Jolicoeur and J. Riera, Phys. Rev. {\bf B42},
6283 (1990).

\item{\bf 15.}
J. D. Reger and A. P. Young, Phys. Rev. {\bf B37}, 5978 (1988).

\item{\bf 16.}
J. Ferrer, Rutgers preprint (1992).

\item{\bf 17.}
R. Chitra, S. Rao and D. Sen, in preparation.

\item{\bf 18.}
C. K. Majumdar and D. K. Ghosh, J. Math. Phys. {\bf 10}, 1388
(1969); C. K. Majumdar, J. Phys. {\bf C3}, 911 (1970).

\item{\bf 19.}
S. Rao and D. Sen, preprint TIFR/TH/92-56, IISc/CTS/92-9 (1992).

\item{\bf 20.}
P. Azaria, B. Delamotte and D. Mouhanna, Phys. Rev. Lett. {\bf 68},
1762 (1992) and Phys. Rev. {\bf B45}, 12612 (1992).

\item{\bf 21.}
J. Villain, J. Phys. (Paris) {\bf 35}, 27 (1974).

\item{\bf 22.}
H. Nishimori and S.J. Miyake, Prog. of Theor. Phys.
{\bf 73}, 18 (1985).

\vfil
\eject
\end